\documentclass[twocolumn]{aastex63}
\usepackage{graphics,epsf}
\usepackage{amsmath}                
\usepackage{amsfonts}               
\usepackage{amssymb}                
\usepackage{epsfig}                 
\usepackage{appendix}
\usepackage{graphicx}
\usepackage{float}
\usepackage{color}
\usepackage{multirow}
\usepackage{colortbl}
\usepackage[para,online,flushleft]{threeparttable}

\newcommand{\yr}{{~\rm yr}}

\newcommand{\AU}{{~\rm AU}}

\newcommand{\days}{{~\rm days}}

\begin{document}

\title{Accretion induced merger leading to core collapse supernovae in old stellar populations}


\author{Jessica Braudo}
\affiliation{Department of Physics, Technion – Israel Institute of Technology, Haifa 3200003, Israel}

\author{Ealeal Bear}
\affiliation{Department of Physics, Technion – Israel Institute of Technology, Haifa 3200003, Israel}

\author{Noam Soker}
\affiliation{Department of Physics, Technion – Israel Institute of Technology, Haifa 3200003, Israel}
\affiliation{Guangdong Technion Israel Institute of Technology, Guangdong Province, Shantou 515069, China}

\email{ealealbh@gmail.com; soker@physics.technion.ac.il}

\begin{abstract}
We examine a triple-star evolution that might lead to core collapse supernovae (CCSNe) in stellar populations that are too old to allow for single or binary evolution to form CCSNe, i.e., where the most massive stars that evolve off the main sequence have masses of $\simeq 4-5 M_\odot$. 
In the scenario we examine the most massive star in the triple system, of mass $\simeq 4-5 M_\odot$, transfers mass to an inner binary system at an orbital separation of $\simeq 100-1000 R_\odot$. The initial orbital separation of the inner binary is $\simeq 10-50 R_\odot$. The inner binary accretes most of the mass that the primary star loses and the two stars expand and their mutual orbit contracts until merger. The merger product is a main sequence star of mass $\simeq 8-10 M_\odot$ that later experiences a CCSN explosion and leaves a NS remnant, bound or unbound to the white dwarf (WD) remnant of the primary star. 
We estimate the event rate of this WD-NS reverse evolution scenario to be a fraction of $\approx 5 \times 10^{-5}$ of all CCSNe. We expect that in the coming decade sky surveys will detect 1-5 such events. 
\end{abstract}

   
\section{Introduction} 
\label{sec:intro}

Singly-evolving stars explode as core collapse supernovae (CCSNe), including electron-capture supernovae (SNe), if their zero-age main sequence (ZAMS) mass is $M_{\rm ZAMS} \ga 8.5-9 M_\odot$. This mass limit depends on some uncertain stellar structure parameters as well as on initial metallicity (e.g., \citealt{Hegeretal2003, Poelarendsetal2008, IbelingHeger2013, Dohertyetal2017, GilPonsetal2018}).   

Binary evolution, through merger and mass transfer, might allow ZAMS stars with lower mass to end as CCSNe and leave neutron star (NS) remnants, e.g., electron-capture SNe by mass transfer (e.g., \citealt{SiessLebreuilly2018}) and by binary merger (e.g., \citealt{Zapartasetal2017}). \cite{Zapartasetal2017} conduct a population synthesis study and find that $\simeq 10-20 \%$ of CCSNe occur by the merging of two stars in the mass range of  $\simeq 4-8 M_\odot$. This merger yields a massive enough star to end as a CCSN.
 
If a star with an initial mass of $M_{\rm ZAMS} \la 8 M_\odot$ explodes as a CCSN, it must be in a multiple stellar system, e.g., a binary or a triple stellar system.
The main routes involve the merger of two stars or mass transfer, both processes bring the CCSN progenitor to a mass of $\ga 8.5 M_\odot$ (or even  $> 8 M_\odot$ might allow CCSN). In that case some of the evolutionary routes form a white dwarf (WD) before the CCSN explosion. This binary evolutionary channel is termed a {\it WD-NS reverse evolution} (e.g., \citealt{SabachSoker2014}). The initial more massive star in the binary system, the primary star, has a mass of $5 M_\odot \la M_{\rm 1,ZAMS} \la 8-8.5 M_\odot$. It evolves first, transfers mass to its companion (the secondary star) and forms the WD. The mass-transfer brings the secondary star to have a post-accretion mass of $M_{\rm 2,B, f} \ga 8.5 M_\odot$, and so it might end its life as a CCSN, leaving behind a NS remnant. If the mass transfer takes place after the secondary star has developed a helium core, i.e., it has already left the main sequence, then the secondary star is not likely to explode as a CCSN even if its mass after mass transfer is $M_{\rm 2,B, f} \ga 8.5 M_\odot$ (e.g., \citealt{BearSoker2021}).
We refer to the initial second massive star as the `secondary star', whether it loses or gains mass. Similarly, we refer to the initial more massive star as the `primary star'. We refer to the initially least massive star in a triple system as the `tertiary star'.  

Accretion induced collapse, where a WD accretes mass and collapses to a NS, and similar processes involve the formation of a WD before a NS 
(e.g. \citealt{Ruiteretal2019, TaurisJanka2019, LiuWang2020, WangLiu2020}).
As we mentioned above, another WD-NS reverse evolution is of a binary system that experiences a mass-transfer process that brings the secondary star to evolve toward a CCSN and leaves behind a WD-NS binary system, bound or unbound 
(e.g.,  \citealt{TutukovYungelSon1993, PortegiesZwartVerbunt1996,  TaurisSennels2000, Brownetal2001, Nelemansetal2001, Kimetal2003, Kalogeraetal2005, vanHaaftenetal2013, Toonenetal2018, Breiviketal2020}). The WD-NS reverse evolution might account for the binary radio pulsars PSR~J1141-6545 and PSR~B2303+46 that have a NS-WD binary system each (e.g., \citealt{PortegiesZwartYungelson1999, vanKerkwijkKulkarni1999,  TaurisSennels2000, Brownetal2001, Daviesetal2002, Churchetal2006}). 
\cite{SabachSoker2014} and \cite{Ablimit2021PASP} discuss other outcomes of the WD-NS reverse evolution, including the merger of the WD with the core of the NS progenitor, and some types of bright transient events (intermediate luminosity optical transients; ILOTs).

There are rare and peculiar exploding stars that require the WD-NS reverse evolution to take place in triple stellar systems. Such is the formation of CCSN inside a planetary nebula \citep{BearSoker2021}. \cite{BearSoker2021} find that in binary systems that experience the WD-NS reverse evolution the minimum time from the formation of planetary nebula and the CCSN explosion is about million years. By the explosion time the planetary nebula has long dispersed into the interstellar medium. Starting with two stars of very close initial masses does not help because in this case the primary transfers mass to the secondary star after the secondary star has left the main sequence and developed a helium core. This late mass transfer implies that even that its mass exceeds $\simeq 8.5 M_\odot$ the secondary star does not explode, but rather forms a planetary nebula and leaves a WD remnant. 
\cite{BearSoker2021} conclude that a CCSN explosion might occur inside a planetary nebula only if a third star forms the PN. This is a rare case because the third star should be less massive than the secondary star but by no more than ${\rm few} \times 0.01 M_\odot$. \cite{BearSoker2021} estimate that about one CCSN inside a planetary nebula occurs out of $10^4$ CCSNe. 

In the present study we examine a WD-NS reverse evolution where the most massive star in the system has an initial (ZAMS) mass of $M_{\rm 1,ZAMS} \simeq 4 - 5 M_\odot$. In a binary system mass transfer will not bring the secondary star to be sufficiently massive to explode as a CCSN. A merger of two stars with combined mass of $\ga 8.5 M_\odot$ might lead to CCSN, but for a CCSN a very fine tuned mass range is needed if at all it can take place, and no WD formation will take place. We consider therefore a triple stellar system where the primary star transfers mass to an inner (tight) binary system, and forms a WD. The mass transfer causes the two stars of the inner binary to merge while they are on the main sequence such that the mass of the inner binary system together with the mass it accreted from the primary star add up to a star of mass $M_{\rm 23acc} \ga 8.5 M_\odot$, that later explodes as a CCSN. 
We base our scenario on the process where the mass transfer causes the inner binary to merge (for a recent thorough study of this process see, e.g., \citealt{DiStefano2020}). 

\cite{deVriesetal2014} already studied the mass transfer from a tertiary star to an inner tight binary system, but not for systems that can experience the WD-NS reverse evolution. They claim that triple-star systems where one star suffers Roche lobe overflow (RLOF) to a tight (inner) binary system amount to $\la 1\%$ of triple-star systems. However, in a newer paper \cite {Leighetal2020} find that $>10\%$ of all triple star systems might have circumbinary disk around the inner binary.  \cite{deVriesetal2014} also find that under their assumptions for the cases they studied the accreted mass does not form an accretion disk around neither of the stars in the inner binary system, nor around the binary (a circumbinary disk). The inner binary system ejects most of the mass, and accretes only about fifth of the mass that the primary star transfers. They used an adiabatic equation of state and did not allow for radiative cooling. In the large orbital separations we use here we expect the gas that leaves the primary star to be optically thin and cool by radiation. This cooling is likely to allow a higher accretion rate. The orbit of the inner binary does shrink. In the two cases that \cite{deVriesetal2014} simulated the initial ratio of the inner semi-major axis (of the inner binary system)  to the triple semi-major axis were $a_{\rm in}/a_{\rm out}=0.046$ and $a_{\rm in}/a_{\rm out}=0.108$. They simulated observed systems. 

In a later study \cite{PortegiesZwartLeigh2019} examine the mass transfer from a low mass star to an inner binary system through an accretion  disk. This leads to the formation of two blue stragglers. They find that the inner binary system accretes at least $80\%$ of the mass that the primary star transfers (they use the term tertiary star for their mass-losing star). We will therefore also assume that the inner binary system accretes a large fraction of the mass that the primary star transfers.   

\cite{PortegiesZwartLeigh2019} mention that the accreting binary system might evolve towards a mass ratio 1 in some cases (the lower mass star in the inner binary system accretes more mass.), but not in all cases.  \cite{deVriesetal2014} find that the more massive star in the inner binary accretes more mass. 
We will assume that the two stars of the inner binary accrete relative to their respective masses. In section \ref{sec:scenario} we describe the proposed scenario.   

We describe the evolution to mass transfer in section \ref{sec:MassTransfer} and the merger of the tight binary system in section \ref{sec:merger}. In section \ref{sec:Rate} we crudely estimate the rate of such very rare events. 
We summarize our study in section \ref{sec:summary}

\section{The proposed scenario}
\label{sec:scenario}

\begin{figure*}
\includegraphics[trim=1.0cm 6.0cm 0.0cm 3.0cm ,clip, scale=0.80]{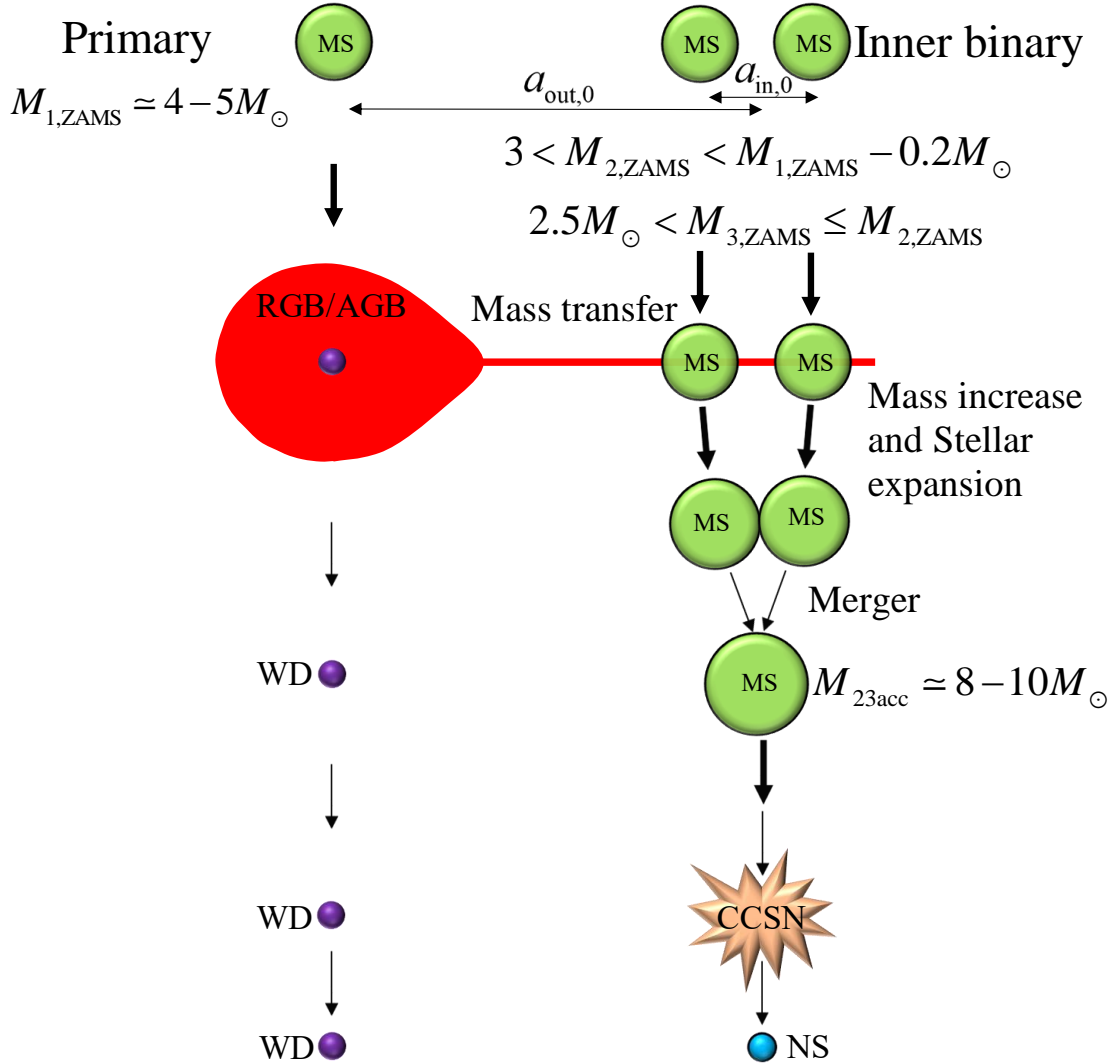}
\caption{A schematic diagram of our proposed WD-NS reverse triple star evolution to a CCSN in an old stellar population where binary mass transfer cannot lead to a CCSN. 
Thick arrows depict phases we do simulate, while thin arrows are phases we do not simulate in this study. 
Abbreviation: AGB: asymptotic giant branch; CCSN: core-collapse supernova;  MS: main sequence; NS: neutron star; RGB: red giant branch. WD: white dwarf.}
 \label{Fig:SchematicScenario}
 \end{figure*}

The main process of the scenario is the merger of two stars with an additional mass from the primary star, such that the mass of the merger remnant is $M_{\rm 23acc} \simeq 8-10 M_\odot$, i.e., sufficiently massive to explode as a CCSN. This supposes to take place in an old enough stellar population where binary merger alone cannot bring a merger remnant to be massive enough to explode as a CCSN. 
We schematically present the proposed scenario in Fig. \ref{Fig:SchematicScenario}. 

We will refer to the initially most massive star as the primary along the entire evolution. Its initial mass is  $M_{\rm 1,ZAMS} \simeq 4 - 5 M_\odot$ and it is on a wide orbit of $a_{\rm out,0} \approx 0.5-5 \AU$ with an inner tight binary system with initial masses of $M_{\rm 2,ZAMS}$ and $M_{\rm 3,ZAMS}$. The primary star evolves to become a giant, and either on its red giant branch (RGB; for $a_{\rm out,0} \la 1 \AU$, depending on eccentricity) or on its asymptotic giant branch (AGB; for $a_{\rm out,0} \ga 1 \AU$) it transfers mass to the inner binary system via RLOF such that the inner binary system accretes most of the mass that the primary loses.

To the condition that the primary mass is not large enough to allow CCSN by mass transfer, there are two additional conditions on the initial masses of the inner binary components.  
($i$) The primary star should expand to become a giant and transfer mass to the tight binary system before the secondary develops a helium core, i.e., before it leaves the main sequence. Otherwise it will not explode later as a CCSN (e.g., \citealt{BearSoker2021}). This condition reads $M_{\rm 2 ,ZAMS}\la M_{\rm 1,ZAMS} - 0.2M_\odot$. 
($ii$) The mass of the merger remnant of the inner binary system with the mass it accretes from the primary star, $M_{\rm acc}$, should form a main sequence star that is massive enough to explode as a CCSN, i.e., $M_{23acc} = M_{\rm 2 ,ZAMS}+M_{\rm 3 ,ZAMS} + M_{\rm acc} \simeq 8 - 10 M_\odot$. 
More massive stars also explode, but in the scenario we study here the mass will not be much larger that $M_{\rm 23acc} \simeq 8-9 M_\odot$. 
This condition reads $2.5 M_\odot \la M_{\rm 3,ZAMS} \le M_{\rm 2,ZAMS}$. 
  
The conditions on the outer and inner orbits are as follows. 
The outer orbit, of the primary star and the inner (tight) binary system, should allow a RLOF mass transfer. A common envelope evolution might also take place, but the scenario does not allow a too large mass ejection as the tight binary system should accrete most of the mass that the primary system loses. A common envelope evolution is likely to eject too much mass. This condition reads therefore $0.5 \la a_{\rm out,0} \la 5 \AU$, strongly depending on the eccentricity. We will not explore this parameter space in the present study because our goal is to demonstrate the possibility of the scenario. We will assume a circular orbit and examine one initial value of $a_{\rm out,0}$. 
                                                              
The condition on the initial inner orbit $a_{\rm in,0}$, that of the inner binary system, is that the two stars merge during (or shortly after) mass transfer.  We study this constrain on $a_{\rm in,0}$ in section \ref{sec:merger}.  

The final outcome of this triple star WD-NS reverse evolution is a CCSN that leaves a NS in a bound or unbound orbit with a WD. As we emphasize, this scenario might take place in relatively old stellar populations where even binary merger at late times cannot lead to CCSNe.   

\section{Evolution to mass transfer}
\label{sec:MassTransfer}

To follow the evolution from the main sequence to the end of the phase of mass transfer from the primary to the inner binary system we use version 10398 of the stellar evolutionary code \textsc{mesa-binary} \citep{Paxtonetal2011,Paxtonetal2013,Paxtonetal2015,Paxtonetal2018,Paxtonetal2019}.

As an example of a triple stellar system that might lead to a CCSN (including the possibility of electron capture SN) we start with three ZAMS stars of 
$M_{\rm 1, ZAMS}=4 M_\odot$ (the primary) on a large orbit, and an inner (tight) binary system of masses $M_{\rm 2, ZAMS}=3.5 M_\odot$ and $M_{\rm 3, ZAMS}=3 M_\odot$.
We consider only circular orbits. We take the orbital separation of the primary star from the center of mass of the inner binary system to be $a_{\rm out,0}=500R_\odot$.  { As we show later, the initial inner orbital separation should be $10 \la a_{\rm in,0} \la 50 R_\odot$. We also assume that the triple system is coplanar, i.e., the inner and outer orbital planes are the same, and the orbits are circular. As both eccentricities and the inclination of the two orbital planes are not large this system is stable to the  Kozai-Lidov mechanism (e.g., \citealt{Katzetal2011, Naozetal2014}). Moreover, tidal forces further stabilize the system against the Kozai-Lidov mechanism (e.g., \citealt{Liuetal}). The system obeys also the dynamical stability criterion for coplanar prograde systems according to equation (90) of \cite{MardlingAarseth2001}, which for the parameters of the systems implies that $a_{\rm in,0} \la 0.3 a_{\rm out,0}$. Namely, the inner binary will not experience merger before mass transfer occurs.  }

In the first part of the simulation we do not follow the evolution of the inner binary system, but rather refer to it as a point mass of $M_{\rm point} = M_{\rm 2, ZAMS}+M_{\rm 3, ZAMS}= 6.5 M_\odot$  (for a recent population synthesis study of mass transfer to an inner binary see \citealt{Hamersetal2021}).  This point mass can accrete mass. We continue the evolution to the end of the mass transfer phase from the primary to the inner binary system. 

As in our earlier paper \cite{BearSoker2021} we base our \textsc{mesa-binary} mode on the \textit{inlists} (i.e., code prescription) of \cite{GibsonStencel2018}. This uses the mass-transfer scheme of \cite{KolbRitter1990} with mass-transfer efficiency scheme from \cite{Sobermanetal1997} (as \citealt{GibsonStencel2018} apply).  
We take the fraction of the mass-loss from the fast wind in the vicinity of the donor star $M_1$ to be $\alpha=0.1$ and the mass-loss fraction from the vicinity of the accretor point mass (that represents the inner binary system) to be $\beta=0.1$. We take $\delta = 0.1$ to be the mass-loss fraction from the circumbinary co-planar toroid. These three identical factors imply that the inner binary accreted $90 \%$ of the mass that the primary loses. 
The radius of this circumbinary toroid is $\gamma^2 a$, where $a$ is the binary semi-major axis and $\gamma = 1.3$ \citep{GibsonStencel2018}. 
The initial metalicity is $Z=0.01$ and the initial equatorial surface rotation velocity of $M_1$ is $v_{\rm ZAMS, 1} = 2~{\rm km}~{\rm s}^{-1}$. 

\begin{figure}
\includegraphics[trim=3.0cm 8.0cm 0.0cm 8.0cm ,clip, scale=0.55]{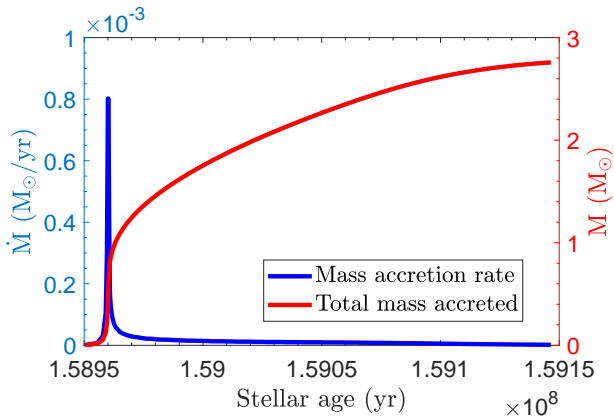}
\caption{Mass accretion rate by the inner binary system (blue line)  and the total accreted mass (red line) as function of time from ZAMS. The total mass the inner binary accretes under our assumptions is $\Delta M_{\rm acc}=2.82 M_\odot$. The graph presents a time period of $2 \times 10^5 \yr$. }
 \label{Fig:MassTransfer}
 \end{figure}

In Fig. \ref{Fig:MassTransfer} we present the mass accretion rate onto the tight binary system as function of time. The mass transfer takes place while the primary star is on the AGB as it has a radius of $164 R_\odot$, has developed a carbon core of $M({\rm C,c})=0.51M_\odot$, and the boundary of the helium core is at $M({\rm He,c})=0.79M_\odot$.  

 At the end of the mass transfer phase the outer orbital separation is $a_{\rm out}=2430 R_\odot$ and the WD mass is $M_{\rm WD}=0.8 M_\odot$.
Such a large orbital separation implies that the merger product will evolve as a single star for most, or even all, of its evolution. In section \ref{sec:merger} we describe a simulation we conduct that shows that the WD in the case we simulate does not influence much the evolution of the merger remnant and does not prevent its final explosion. 
  
\section{Merger of the inner binary system}
\label{sec:merger}
\begin{figure}
\includegraphics[trim=3.0cm 8.0cm 0.0cm 8.0cm ,clip, scale=0.55]{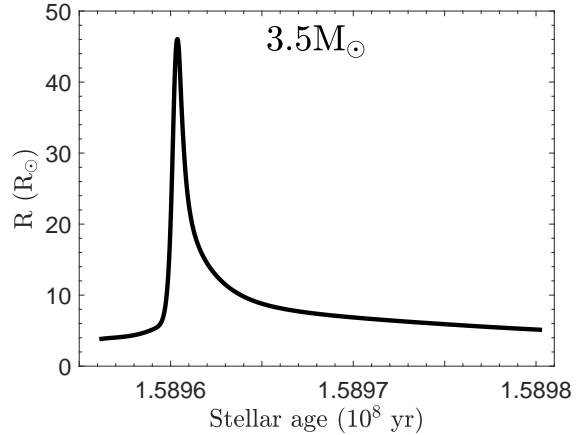} \\
\includegraphics[trim=3.0cm 8.0cm 0.0cm 8.0cm ,clip, scale=0.55]{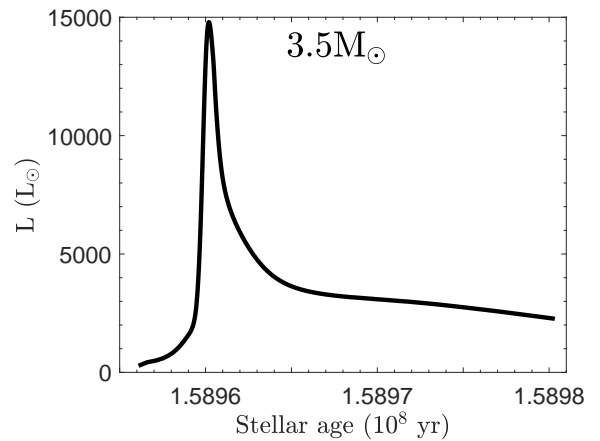}
\caption{The evolution of the radius (upper panel) and luminosity (lower panel) of the $M_{\rm ZAMS,2}=3.5 M_\odot$ component of the inner binary during the main mass accretion phase. The total accreted mass by this star under our assumptions is $\Delta M_{\rm acc,2}=1.52 M_\odot$. The time between two consecutive small ticks is $1000 \yr$.  }
 \label{Fig:RLstar35}
 \end{figure}

\begin{figure}
\includegraphics[trim=3.0cm 8.0cm 0.0cm 8.0cm ,clip, scale=0.55]{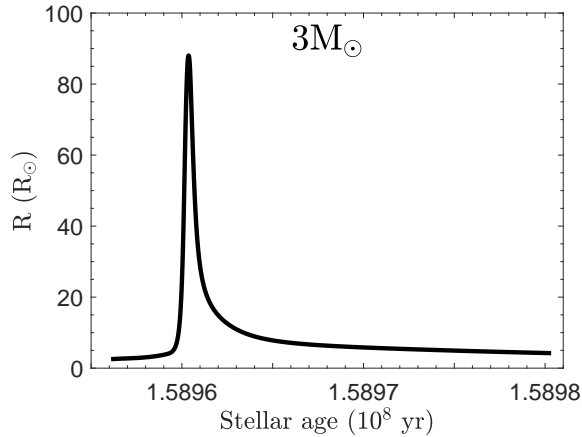} \\
\includegraphics[trim=3.0cm 8.0cm 0.0cm 8.0cm ,clip, scale=0.55]{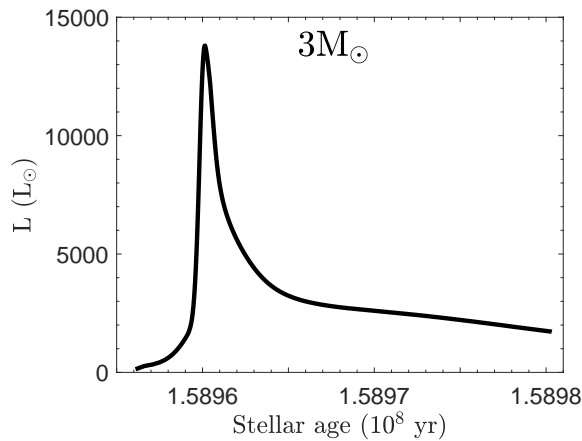}
\caption{Similar to Fig. \ref{Fig:RLstar35} but for the $M_{\rm ZAMS,3}=3 M_\odot$ component of the inner binary.
}
 \label{Fig:RLstar30}
 \end{figure}

We follow the response of the two stars of the inner (tight) binary system to the mass they accrete. Mass accretion by the binary system is likely to cause orbital contraction (e.g., \citealt{Comerfordetal2019}). We assume that the mass that each star accretes is proportional to its mass. The secondary and the tertiary stars accreted each a total mass of $\Delta M_{\rm 2,acc}=1.52 M_\odot$ and $\Delta M_{\rm 3,acc}=1.3 M_\odot$, respectively. In Figs. \ref{Fig:RLstar35} and \ref{Fig:RLstar30} we present the evolution of the radius and luminosity of each of these two stars during the main mass transfer phases. 

At the peak of mass transfer rate the mass accretion time scale $M/\dot M_{\rm acc} \simeq 10^4 \yr$ is much shorter than the thermal time scale of each of the two stars $\tau_{\rm KH} > 10^5 \yr$. This explains the very large expansion of the stars.  

The orbital separation of the inner binary system changes as a result of mass accretion in a way that depends on the angular momentum of the accreted mass. For a zero angular momentum of the accreted mass, $J_{\rm acc}=0$, angular momentum conservation gives the final orbital separation as 
\begin{equation}
a_{\rm in} (J_{\rm acc}=0)= a_{\rm in,0} 
\left( \frac{M_{\rm 2,f}+M_{\rm 3,f}}{M_{\rm 2,b}+M_{\rm 3,b}} \right)
\left( \frac{M_{\rm 2,b} M_{\rm 3,b}}{M_{\rm 2,f} M_{\rm 3,f}} \right)^2,
\label{eq:a23i}
\end{equation}
where subscript $b$ stands for just before mass accretion starts and $f$ stands for after mass transfer ends. We take the stellar masses just before mass transfer starts to be very similar to their ZAMS masses as winds on the main sequence do not carry much mass. Since we distribute the accreted mass by the initial mass of the stars we can write equation (\ref{eq:a23i}) as 
\begin{equation}
a_{\rm in} (J_{\rm acc}=0) = a_{\rm in,0} 
\left( \frac{M_{\rm 2,b} + M_{\rm 3,b}}
{M_{\rm 2,b} + M_{\rm 3,b} + \Delta M_{\rm acc}} \right)^3.
\label{eq:a23ii}
\end{equation}

The accreted mass has some angular momentum. If the angular momentum of the inner binary system and the triple system have the same sense, so will the accreted mass and the final orbital separation will be larger than what equation (\ref{eq:a23ii}) gives. However, the angular momenta of the two stars might also be misaligned, or even have opposite sense such that the final orbital separation is smaller than what equation (\ref{eq:a23ii}) gives. In checking whether the mass transfer by itself can induce merger we will use equation (\ref{eq:a23ii}) as it is.
Overall, we take as our merger condition by mass accretion that at the end of mass transfer 
 \begin{equation}
R_{\rm 2} + R_{\rm 3} \ge a_{\rm in} (J_{\rm acc}=0). 
\label{eq:Merger}
\end{equation}

\begin{figure}
\includegraphics[trim=3.0cm 8.0cm 0.0cm 8.0cm ,clip, scale=0.55]{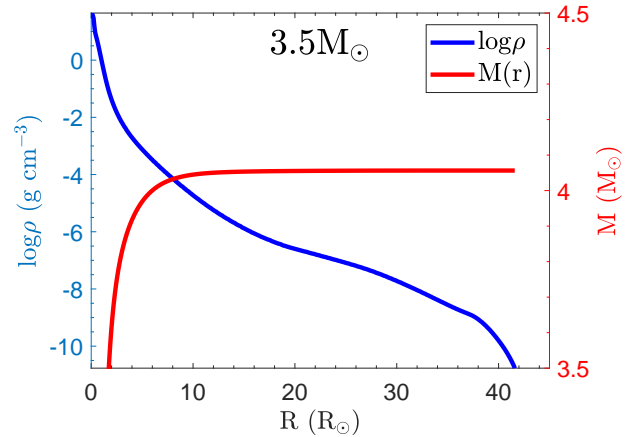} \\
\includegraphics[trim=3.0cm 8.0cm 0.0cm 8.0cm ,clip, scale=0.55]{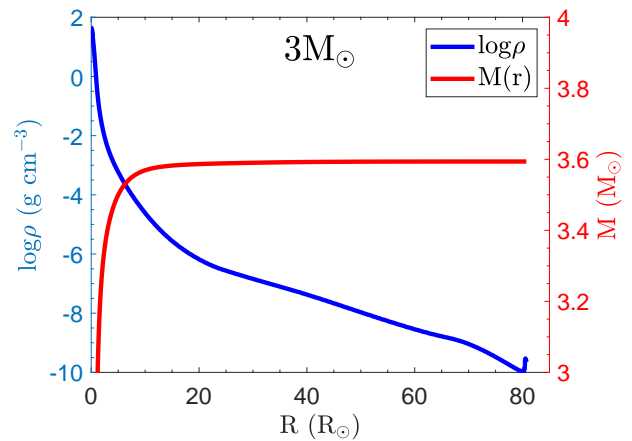}
\caption{The density profiles (blue lines) and the accreted mass profile (red lines) of the two stars of the binary system at their maximum expansion. 
}
 \label{Fig:Densities}
 \end{figure}

The evolution of the stars (Figs. \ref{Fig:RLstar35} and \ref{Fig:RLstar30}) show that the large expansion takes place during the high mass transfer rate, before much mass has been accreted. To apply condition (\ref{eq:Merger}) after the inner binary accreted most of the mass, i.e., $\Delta M_{\rm acc} = 2.8 M_\odot$, would require using $R_{\rm 2} + R_{\rm 3} \simeq 8 R_\odot$. With this left hand side of equation (\ref{eq:Merger}), the merger condition reads
$a_{\rm in,0} \la 24 R_\odot$. However, the two stars expand to huge dimensions. This does not mean that we can  place them at much large separations as the inflated envelopes are of very low density as we show in Fig. \ref{Fig:Densities}.  

At maximum expansion $(R_2+R_3)_{\rm max}=130R_\odot$. However, due to the very low masses that reside in the very outer zones of the two inflated envelopes, we crudely take the maximum orbital separation for merger to be $\approx 50 R_\odot$. 
To prevent merger before mass transfer (so the binary merger does not evolve fast and off the main sequence) we require the initial orbital separation to be $a_{\rm in,0} \ga 10 R_\odot$. This allowed range of initial orbital separation, 
$10 \la a_{\rm in,0} \la 50 R_\odot$,
corresponds to an allowed period interval of $ 1~{\rm day} \la P_{\rm in,0} \la 16 {\rm day}$.

 We cannot simulate the full merger process. Instead, we evolved a star with an initial mass of $M_{\rm 2,ZAMS} = 3.5 M_\odot$ to the time when it merges with the $M_{\rm 3,ZAMS} = 3 M_\odot$ after they have accreted a mass of $\Delta M_{\rm acc}=2.82 M_\odot$.
We added a mass of $3M_\odot + 2.82M_\odot$ of solar composition to the evolved $M_{\rm ZAMS} = 3.5 M_\odot$ star to form a new star of mass $M_{\rm 23acc} =  9.32 M_\odot$. 
We evolved this simulated merger product with a WD companion of mass $M_{\rm WD}=0.8 M_\odot$. (a point mass) starting with an orbital separation of $a=a_{\rm out}=2430 R_\odot$ using \textsc{mesa-binary}. Because of numerical difficulties we could not follow the evolution to the point of electron capture that leads to explosion. We had to terminate the evolution when the ONeMg core reached a mass of $M_{\rm core}=1.2 M_\odot$. The fate of such stars, i.e., in the mass range we focus on here $M_{\rm ZAMS} \approx 9 M_\odot$, is determined by the competition between core growth and stellar mass loss (e.g., discussion by \citealt{Poelarendsetal2017}). We find that at the termination time of the evolution of the remnant product the ratio of the stellar mass loss rate to core-mass growth rate is about 10. The envelope mass at the time we terminated the evolution is $M_{\rm env}=7.4 M_\odot$. By the time the core will reach a mass of $1.4 M_\odot$ the mass loss would remove only about $2 M_\odot$ from the envelope.  There are uncertainties in the final fate of the star because the competition between core growth and mass loss depends on some uncertain processes, such as the amount of overshooting introduced into the model and the stellar mass loss rate. However, since for the present parameters our stellar model will lose less than third of the envelope mass before it reaches the minimum core mass for explosion $\simeq 1.4 M_\odot$, there is a large margin for the core to reach this minimum mass. We also find that the orbital separation of the WD increases according to the mass loss rate. Namely, tidal interaction is not strong enough to shrink the orbit of the WD. This implies that the WD will not influence the fate of the merger product to explode. 

From the above discussion we conclude that with a large margin, the core will reach the mass of electron capture long before the star loses its envelope. The star will explode as a type II CCSN (electron capture) as the hydrogen-rich envelope mass will be non-negligible. 
Furthermore, \cite{Poelarendsetal2008} find that stars that reach a maximum luminosity of $L > 10^5 L_\odot$ in the above mass range explode as electron capture SNe.  Our model of the merger remnant when we terminate its evolution has a luminosity of $1.2 \times 10^5 L_\odot$. We safely conclude that the merger remnant of our scenario will explode as an electron capture SN at $t=1.77 \times 10^8 \yr$ after star formation.   

The main conclusion from this section is that there is a non-negligible volume of the parameter space that allows the triple star WD-NS reverse evolution that we study here to take place. 

\section{Rate}
\label{sec:Rate}

We crudely estimate the rate of CCSNe from triple star evolution in populations that are too old for binary merger or binary mass transfer to form CCSNe. Namely, populations where stars of $4 M_\odot < M_{\rm 1,ZAMS} <5 M_\odot$ are now evolving off-the main sequence. We find the following factors. 

(1) For an initial mass function of $dN \propto M^{-2.35}$ (e.g. \citealt{Bastianetal2010}) the fraction of primary stars with $4 M_\odot < M_{\rm 1,ZAMS} <5 M_\odot$ to the number of CCSNe, which we take as all stars with mass $>8 M_\odot$, is $f_{M1} \simeq 0.7$.  

(2) The probability of such stars to be in triple or higher order stellar systems if $f_{\rm T} \simeq 0.3$ (e.g., \citealt{MoeDiStefano2017}). 

(3) From the mass ratio distributions that \cite{MoeDiStefano2017} give for these primary masses, we find that the probability of the secondary star to be of mass $M_{\rm 2,ZAMS} \simeq 0.8-0.95 M_{\rm 1,ZAMS}$ is $f_2 \simeq 0.15$, with a similar ratio for the tertiary star $f_3 \simeq 0.15$. 

(4) From \cite{MoeDiStefano2017} we find that the fraction of systems with an outer orbital semi-major axis of $a_{\rm out,0} \simeq 100-1000 R_\odot$, i.e., an orbital period range of $P_{\rm out,0} \simeq 30 - 1000 \days$, is $f_{\rm out} \simeq 0.15$. 

(5) In section \ref{sec:merger} we found that the initial orbital period range for the inner binary is $ 1~{\rm day} \la P_{\rm in,0} \la 16 {\rm day}$. From \cite{MoeDiStefano2017} we find the probability for this range to be $f_{\rm in} \simeq 0.07$.   

Over all, the fraction of CCSNe in old stellar population where only triple star mass transfer can lead to CCSNe to the total CCSNe rate is 
\begin{equation}
F_{\rm Old~Triple} \simeq f_{M1} f_{\rm T} f_2 f_3 f_{\rm out} f_{\rm in}  \approx 5 \times 10^{-5}. 
\label{eq:Foldtriple}
\end{equation} 

As expected, old-triple CCSNe are very rare. However, we expect some to be observed. With new sky transient surveys with an expected CCSN detection rate of $\approx 10^4 \yr^{-1}$, we can expect that within the next decade one to few CCSNe will be observed in old stellar populations where a binary mass transfer cannot account for these CCSNe, but triple star evolution can. 

\section{Summary}
\label{sec:summary}

Sky surveys, e.g., the All-Sky Automated Survey for Supernovae (ASAS-SN; \citealt{Kochaneketal2017PASP}), the Zwicky Transient Facility (ZTF; \citealt{Bellmetal2019}), the Southern Hemisphere Variability Survey (LSQ; \citealt{Baltayetal2013}), and the Large Synoptic Survey Telescope (LSST; \citealt{Ivezicetal2019}), will detect $\approx 10^4$ CCSNe per year. This number implies that during a time period of several years these surveys will observe even very rare types of CCSNe and CCSN impostors. This motivates us to conduct a series of studies of a variety of rare CCSNe and impostors (e.g., \citealt{BearSoker2021SNIaCCSNE, Schreieretal2021}). The present study is another study in that series of papers, now focusing on CCSNe in stellar populations that are too old for binary mass transfer to result in a CCSN. We rather consider a triple-star evolution that might lead to CCSNe in stellar populations where the most massive stars that evolve off the main sequence have masses of $M_{\rm 1, ZAMS} \simeq 4-5 M_\odot$ (Fig.   \ref{Fig:SchematicScenario}).  As the evolution forms a WD before it forms the NS remnant of the CCSN, it is a WD-NS reverse evolution. 

In the present study we do not search the parameter space for this scenario because the uncertainties are too large, e.g., how much mass the inner binary accretes from the mass the primary loses. We demonstrated the scenario for one set of parameters by examining the mass transfer rate as function of time (section \ref{sec:MassTransfer}; Fig. \ref{Fig:MassTransfer}), and then by following the evolution of each of the two stars of the inner binary system as they accrete mass (section \ref{sec:merger}; Figs. \ref{Fig:RLstar35}-\ref{Fig:Densities}).  For the parameters we use here explosion takes place at $t=1.77 \times 10^8 \yr$ after star formation.  We then estimated the range of initial orbital separations for which the inner binary might merge as a result of the mass transfer (equation \ref{eq:Merger}).  

In section \ref{sec:Rate} we crudely estimated the event rate of the scenario we studied here to be a fraction $F_{\rm Old~Triple} \approx 5 \times 10^{-5}$ of CCSNe. Although very rare, as we mentioned above, we expect that in the coming decade sky surveys will detect 1-5 such events. 
 
The results of this study add to the call to consider rare triple-star evolutionary routes when analysing observations of peculiar and very rare  CCSNe and CCSN impostors.

\section*{Acknowledgments}
We thank Evgeni Grishin, Erez Michaely, and an anonymous referee for helpful comments. 

This research was supported by a grant from the Israel Science Foundation (769/20).

\textbf{Data availability}

The data underlying this article will be shared on reasonable request to the corresponding author. 


\pagebreak


\begin{thebibliography}{}

\bibitem[Ablimit(2021)]{Ablimit2021PASP} Ablimit, I.\ 2021, \pasp, 133, 074201. doi:10.1088/1538-3873/ac025c

\bibitem[Baltay et al.(2013)]{Baltayetal2013}  Baltay, C., Rabinowitz, D., Hadjiyska, E., Walker, E.~S., Nugent, P., Coppi, P., Ellman, N.,  et al.\ 2013, \pasp, 125, 683. doi:10.1086/671198 
 
\bibitem[Bastian et al.(2010)]{Bastianetal2010} Bastian, N., Covey, K.~R., \& Meyer, M.~R.\ 2010, \araa, 48, 339. doi:10.1146/annurev-astro-082708-101642

\bibitem[Bear \& Soker(2021a)]{BearSoker2021} Bear, E. \& Soker, N.\ 2021a, \mnras, 500, 2850. doi:10.1093/mnras/staa3475

\bibitem[Bear \& Soker(2021b)]{BearSoker2021SNIaCCSNE} Bear, E. \& Soker, N.\ 2021b, \mnras, 506, 919. doi:10.1093/mnras/stab1694

\bibitem[Bellm et al.(2019)]{Bellmetal2019} Bellm, E.~C., Kulkarni, S.~R., Graham, M.~J., Dekany, R., Smith, R.~M., Riddle, R., Masci, F.~J., et al.\ 2019, \pasp, 131, 018002. doi:10.1088/1538-3873/aaecbe 

\bibitem[Breivik et al.(2020)]{Breiviketal2020}  Breivik, K., Coughlin, S., Zevin, M., et al.\ 2020, \apj, 898, 71 


\bibitem[Brown et al.(2001)]{Brownetal2001} Brown, G.~E., Lee, C.-H., Portegies Zwart, S.~F., \& Bethe, H.~A.\ 2001, \apj, 547, 345


\bibitem[Church et al.(2006)]{Churchetal2006} Church, R.~P., Bush, S.~J., Tout, C.~A., et al.\ 2006, \mnras, 372, 715

\bibitem[Comerford et al.(2019)]{Comerfordetal2019}  Comerford, T.~A.~F., Izzard, R.~G., Booth, R.~A., \& Rosotti, G..\ 2019, \mnras, 490, 5196. doi:10.1093/mnras/stz2977 

\bibitem[Davies et al.(2002)]{Daviesetal2002} Davies, M.~B., Ritter, H., \& King, A.\ 2002, \mnras, 335, 369

\bibitem[de Vries et al.(2014)]{deVriesetal2014} de Vries, N., Portegies Zwart, S., \& Figueira, J.\ 2014, \mnras, 438, 1909. doi:10.1093/mnras/stt1688

 \bibitem[Di Stefano(2020)]{DiStefano2020} Di Stefano, R.\ 2020, \mnras, 493, 1855. doi:10.1093/mnras/staa220

\bibitem[Doherty et al.(2017)]{Dohertyetal2017} Doherty, C.~L., Gil-Pons, P., Siess, L., \& Lattanzio, J.~C.,\ 2017, \pasa, 34, e056. doi:10.1017/pasa.2017.52

\bibitem[Gibson \& Stencel(2018)]{GibsonStencel2018} Gibson, J.~L., \& Stencel, R.~E.\ 2018, \mnras, 476, 5026

\bibitem[Gil-Pons et al.(2018)]{GilPonsetal2018} Gil-Pons, P., Doherty, C.~L., Guti{\'e}rrez, J.~L.,Siess, L., Campbell, S.~W., Lau, H.~B., Lattanzio J.~C.,\ 2018, \pasa, 35, 38. doi:10.1017/pasa.2018.42

\bibitem[\protect\citeauthoryear{Hamers, Glanz, \& Neunteufel}{2021}]{Hamersetal2021} Hamers A.~S., Glanz H., Neunteufel P., 2021, arXiv, arXiv:2110.00024

\bibitem[Heger et al.(2003)]{Hegeretal2003} Heger, A., Fryer, C.~L., Woosley, S.~E., Langer, N., \& Hartmann, D.~H.,\ 2003, \apj, 591, 288. doi:10.1086/375341

\bibitem[Ibeling \& Heger(2013)]{IbelingHeger2013} Ibeling, D. \& Heger, A.\ 2013, \apjl, 765, L43. doi:10.1088/2041-8205/765/2/L43

\bibitem[Ivezi{\'c} et al.(2019)]{Ivezicetal2019} Ivezi{\'c}, {\v{Z}}., Kahn, S.~M., Tyson, J.~A., Abel, B., Acosta, E., Allsman, R., Alonso, D., et al.\ 2019, \apj, 873, 111. doi:10.3847/1538-4357/ab042c 

\bibitem[Kalogera et al.(2005)]{Kalogeraetal2005} Kalogera, V., Kim, C., Lorimer, D.~R., et al.\ 2005, Binary Radio Pulsars, 261


\bibitem[\protect\citeauthoryear{Katz, Dong, \& Malhotra}{2011}]{Katzetal2011}  Katz B., Dong S., Malhotra R., 2011, PhRvL, 107, 181101. doi:10.1103/PhysRevLett.107.181101 

\bibitem[Kim et al.(2003)]{Kimetal2003} Kim, C., Kalogera, V., \& Lorimer, D.~R.\ 2003, \apj, 584, 985

\bibitem[Kochanek et al.(2017)]{Kochaneketal2017PASP}  Kochanek, C.~S., Shappee, B.~J., Stanek, K.~Z.,  Holoien, T.~W.-S., Thompson, T.~A., Prieto, J.~L., Dong S., et al.\ 2017, \pasp, 129, 104502. doi:10.1088/1538-3873/aa80d9

\bibitem[Kolb \& Ritter(1990)]{KolbRitter1990} Kolb, U., \& Ritter, H.\ 1990, \aap, 236, 385

\bibitem[Leigh et al.(2020)]{Leighetal2020} Leigh, N.~W.~C., Toonen, S., Portegies Zwart, S.~F., \& Perna, R., \ 2020, \mnras, 496, 1819. doi:10.1093/mnras/staa1670

\bibitem[\protect\citeauthoryear{Li et al.}{2014}]{Naozetal2014}  Li G., Naoz S., Kocsis B., Loeb A., 2014, ApJ, 785, 116. doi:10.1088/0004-637X/785/2/116 

\bibitem[Liu et al.(2015)]{Liuetal}  Liu, B., Mu{\~n}oz, D.~J., \& Lai, D.\ 2015, \mnras, 447, 747. doi:10.1093/mnras/stu2396 

\bibitem[Liu \& Wang(2020)]{LiuWang2020}  Liu, D., \& Wang, B.\ 2020, \mnras, 494, 3422 

\bibitem[\protect\citeauthoryear{Mardling \& Aarseth}{2001}]{MardlingAarseth2001}  Mardling R.~A., Aarseth S.~J., 2001, MNRAS, 321, 398. doi:10.1046/j.1365-8711.2001.03974.x 

\bibitem[Moe \& Di Stefano(2017)]{MoeDiStefano2017} Moe, M. \& Di Stefano, R.\ 2017, \apjs, 230, 15. doi:10.3847/1538-4365/aa6fb6

\bibitem[Nelemans et al.(2001)]{Nelemansetal2001} Nelemans, G., Yungelson, L.~R., \& Portegies Zwart, S.~F.\ 2001, \aap, 375, 890

\bibitem[Paxton et al.(2011)]{Paxtonetal2011} Paxton, B., Bildsten, L., Dotter, A., Herwig, F.,Lesaffre, P. and Timmes, F.\ 2011, \apjs, 192, 3 

\bibitem[Paxton et al.(2013)]{Paxtonetal2013} Paxton, B., Cantiello, M., Arras, P., et al.\ 2013, \apjs, 208, 4 

\bibitem[Paxton et al.(2015)]{Paxtonetal2015} Paxton, B., Marchant, P., Schwab, J., et al.\ 2015, \apjs, 220, 15 

\bibitem[Paxton et al.(2018)]{Paxtonetal2018} Paxton, B., Schwab, J., Bauer, E.~B., et al.\ 2018, The Astrophysical Journal Supplement Series, 234, 34.

\bibitem[Paxton et al.(2019)]{Paxtonetal2019} Paxton, B., Smolec, R., Schwab, J., et al.\ 2019, \apjs, 243, 10, 

\bibitem[Poelarends et al.(2008)]{Poelarendsetal2008} Poelarends, A.~J.~T., Herwig, F., Langer, N.,  \& Heger, A., \ 2008, \apj, 675, 614. doi:10.1086/520872

\bibitem[\protect\citeauthoryear{Poelarends et al.}{2017}]{Poelarendsetal2017}  Poelarends A.~J.~T., Wurtz S., Tarka J., Cole Adams L., Hills S.~T., 2017, ApJ, 850, 197. doi:10.3847/1538-4357/aa988a 

\bibitem[Portegies Zwart \& Leigh(2019)]{PortegiesZwartLeigh2019} Portegies Zwart, S. \& Leigh, N.~W.~C.\ 2019, \apjl, 876, L33. doi:10.3847/2041-8213/ab1b75

\bibitem[Portegies Zwart \& Verbunt(1996)]{PortegiesZwartVerbunt1996} Portegies Zwart, S.~F., \& Verbunt, F.\ 1996, \aap, 309, 179

\bibitem[Portegies Zwart \& Yungelson(1999)]{PortegiesZwartYungelson1999} Portegies Zwart, S.~F., \& Yungelson, L.~R.\ 1999, \mnras, 309, 26

\bibitem[Ruiter et al.(2019)]{Ruiteretal2019}  Ruiter, A.~J., Ferrario, L., Belczynski, K., Seitenzahl, I.~R., Crocker, R.~M., \& Karakas, A.~I.,\ 2019, \mnras, 484, 698 

\bibitem[Sabach \& Soker(2014)]{SabachSoker2014} Sabach, E., \& Soker, N.\ 2014, \mnras, 439, 954

\bibitem[\protect\citeauthoryear{Schreier et al.}{2021}]{Schreieretal2021} Schreier R., Hillel S., Shiber S., Soker N., 2021, MNRAS, 508, 2386. doi:10.1093/mnras/stab2687

\bibitem[Soberman et al.(1997)]{Sobermanetal1997} Soberman, G.~E., Phinney, E.~S., \& van den Heuvel, E.~P.~J.\ 1997, \aap, 327, 620

\bibitem[Siess \& Lebreuilly(2018)]{SiessLebreuilly2018} Siess, L. \& Lebreuilly, U.\ 2018, \aap, 614, A99. doi:10.1051/0004-6361/201732502

\bibitem[Tauris \& Janka(2019)]{TaurisJanka2019}  Tauris, T.~M., \& Janka, H.-T.\ 2019, \apjl, 886, L20 
  
\bibitem[Tauris \& Sennels(2000)]{TaurisSennels2000} Tauris, T.~M., \& Sennels, T.\ 2000, \aap, 355, 236

\bibitem[Toonen et al.(2018)]{Toonenetal2018}  Toonen, S., Perets, H.~B., Igoshev, A.~P., et al.\ 2018, \aap, 619, A53 

\bibitem[Tutukov \& Yungelson(1993)]{TutukovYungelSon1993} Tutukov, A.~V., \& Yungelson, L.~R.\ 1993, Astronomy Reports, 37, 411
  
\bibitem[van Haaften et al.(2013)]{vanHaaftenetal2013} van Haaften, L.~M., Nelemans, G., Voss, R., et al.\ 2013, \aap, 552, A69

\bibitem[van Kerkwijk \& Kulkarni(1999)]{vanKerkwijkKulkarni1999} van Kerkwijk, M.~H., \& Kulkarni, S.~R.\ 1999, \apjl, 516, L25

\bibitem[\protect\citeauthoryear{Wang \& Liu}{2020}]{WangLiu2020} Wang B., Liu D., 2020, RAA, 20, 135. doi:10.1088/1674-4527/20/9/135

\bibitem[Zapartas et al.(2017)]{Zapartasetal2017} E. Zapartas, S. E. de Mink, R. G. Izzard, et al., \aap, 601, A29 (2017) 

\end{thebibliography}
\end{document}